# Topological Insulator Nanowires and Nanoribbons

Desheng Kong,† Jason C. Randel,§,| Hailin Peng,†,⊥ Judy J. Cha,† Stefan Meister,† Keji Lai,‡,§

Yulin Chen,‡,§,| Zhi-Xun Shen,‡,§,| Hari C. Manoharan,‡,| and Yi Cui†,*

*† Department of Materials Science and Engineering, ‡Department of Physics, §Department of Applied Physics, Stanford University, Stanford, California 94305, and |Stanford Institute for Materials and Energy Sciences, SLAC National Accelerator Laboratory, 2575 Sand Hill Road, Menlo Park, California 94025*

Recent theoretical calculations and photoemission spectroscopy measurements on the bulk $Bi_2Se_3$ material show that it is a three-dimensional topological insulator possessing conductive surface states with nondegenerate spins, attractive for dissipationless electronics and spintronics applications. Nanoscale topological insulator materials have a large surface-to-volume ratio that can manifest the conductive surface states and are promising candidates for devices. Here we report the synthesis and characterization of high quality single crystalline $Bi_2Se_3$ nanomaterials with a variety of morphologies. The synthesis of $Bi_2Se_3$ nanowires and nanoribbons employs Au-catalyzed vapor-liquid-solid (VLS) mechanism. Nanowires, which exhibit rough surfaces, are formed by stacking nanoplatelets along the axial direction of the wires. Nanoribbons are grown along [11-20] direction with a rectangular cross-section and have diverse morphologies, including quasi-one-dimensional, sheetlike, zigzag and sawtooth shapes. Scanning tunneling microscopy (STM) studies on nanoribbons show atomically smooth surfaces with ∼ 1 nm step edges, indicating single Se-Bi-Se-Bi-Se quintuple layers. STM measurements reveal a honeycomb atomic lattice, suggesting that the STM tip couples not only to the top Se atomic layer, but also to the Bi atomic layer underneath, which opens up the possibility to investigate the contribution of different atomic orbitals to the topological surface states. Transport measurements of a single nanoribbon device (four terminal resistance and Hall resistance) show great promise for nanoribbons as candidates to study topological surface states.

$Bi_2Se_3$ is a narrow gap semiconductor, previously studied for infrared detectors and thermoelectric applications.[1] Recently, research on $Bi_2Se_3$ and related compounds ($Bi_2Te_3$ and $Sb_2Te_3$) has attracted much interest because they are predicted to be three-dimensional (3D) topological insulators (TIs), a new class of quantum matter possessing conducting surface states with nondegenerate spins.[2] In TIs, the strong spin−orbit coupling dictates robust, nontrivial surface states, which are topologically protected against back scattering from time-reversal invariant defects and impurities. Angle-resolved photoemission spectroscopy (ARPES) measurements on bulk single crystals of $Bi_xSb_{1-x}$, $Bi_2Se_3$, and $Bi_2Te_3$ have verified the existence of the 3D TI phase.[3-7] In particular, the surface states of $Bi_2Se_3$ forms a single Dirac cone inside a large bulk band gap of 0.3 eV, thus being suggested as the reference material for the 3D TIs.[2,4,8] The unique properties of the $Bi_2Se_3$ TI may pave the way for dissipationless quantum electronics and room temperature spintronics applications.

To date, the surface properties of the 3D TIs have been mainly investigated by ARPES measurements on the cleaved surface of bulk crystals.[3-7] Single crystalline nanostructure, on the other hand, offers an attractive alternative system to study the surface states, for example, by transport measurements or surface scanning techniques. As the sample size shrinks, the contribution of surface states to electrical and optical properties is enhanced due to the increase of the surface-to-volume ratio. TI nanostructures are thus an excellent system to investigate the properties of exotic surface states. Also, the study of mesoscopic transport properties is essential for wide applications of TIs in future spintronics devices. In previous reports, $Bi_2Se_3$ nanobelts have been synthesized by sonochemical methods,[9] $Bi_2Se_3$ nanoplates by solvothermal synthesis,[10] and metal−organic chemical vapor deposition.[11] However, surface contaminations due to organic residues or low-crystalline quality by low-temperature solution synthesis are associated with these methods, which may mask the surface states of the TI phase. On the other hand, vapor−liquid−solid (VLS) growth has been shown to be an effective route to produce high quality semiconductor nanomaterials.[12-20] Here we develop Au-catalyzed VLS synthesis of high quality $Bi_2Se_3$ nanowires and nanoribbons. We achieve $Bi_2Se_3$ nanomaterials with high crystallinity and diverse morphologies for surface properties study and device applications. Our scanning tunneling microscopy (STM) study shows atomically flat surfaces of nanoribbons and step edges, which correspond to the thickness of single quintuple layers. These nanoribbons allow for patterning various devices, including the Hall bar configuration, to study key fundamental transport properties.

$Bi_2Se_3$ exhibits a layered, rhombohedral crystal structure in space group $D^5_{3d}$ ($R\bar{3}m$), as shown in Figure 1A. Each charge-neutral layer consists of five covalently bonded atomic planes, Se−Bi−Se−Bi−Se, known as a quintuple layer. These quintuple layers are bonded together predominantly via van der Waals interaction to form the $Bi_2Se_3$ crystal. The conventional unit cell spans over three quintuple layers with lattice parameters $a \sim 4.14$ Å and $c \sim 28.64$ Å.[21] Therefore, each quintuple layer has a thickness of $\sim 1$ nm.

* Corresponding author. E-mail: yicui@stanford.edu.
⊥ Present address: College of Chemistry and Molecular Engineering, Peking University, Beijing 100871, People's Republic of China.

Our group has previously synthesized nanowires and nanoribbons of layer-structured $In_2Se_3$ and GaSe via VLS method.[18-20] Our synthesis of $Bi_2Se_3$ nanomaterials employs a similar growth process and performed in a horizontal tube furnace (Lindberg/Blue M), equipped with a 12 in. length quartz tube. The temperature profile of the furnace is reported previously.[18] The source material, $Bi_2Se_3$ powder (99.999%, 0.2 g per growth from Alfa Aesar), is placed in the hot center of the furnace while substrates Si <100>, functionalized with 0.1% w/v aqueous poly-L-lysine solution (Ted Pella) then coated with 20 nm Au nanoparticles (Sigma Aldrich) are placed in the downstream side of the furnace at distances ranging between 6 to 12 cm from the center. In a typical synthesis process, the tube is initially pumped down to a base pressure of 100 mTorr and flushed with Ar gas several times to remove $O_2$ residue. Then the furnace is heated to high temperature in the range of 450−580 °C (hot center of the tube) and is kept at the high temperature for 1− 5 h under 30 standard cubic centimeters per minute (sccm) Ar carrier gas flow, followed by natural cool-down. A gray coating layer, usually covering the substrate over 1 cm² area along the flow direction, is readily observable by unaided eye.

To understand the morphology of as-grown nanomaterials, we use a FEI XL30 Sirion scanning electron microscope (SEM) to characterize synthesized products (Figure 1). Nanomaterial morphologies depend crucially on the substrates location. At the warm zone (∼8 cm away from the center of the furnace), the growth of quasi-one-dimensional materials (nanowires and narrow ribbons) is dominant (Figure 1B). At the cool zone (∼12 cm from the center), nanoribbons of diverse morphologies with lateral dimensions of several micrometers are the dominant growth product, as shown in Figure 1C. The growth rate of synthesized products increases dramatically with the temperature and flow rate. A representative SEM image of higher temperature growth, for example, at 580 °C with 120 sccm Ar flow for 5 h, is shown in Figure 1D. Large sheets with lateral dimensions of 50− 100 $\mu$m are obtained. Individual nanomaterials with representative morphologies are shown in Figure 1E−J. The presence of a gold nanoparticle at the end of each nanowire and nanoribbon suggests the VLS growth mechanism, in which the gold catalyst induces the nucleation and growth of the nanomaterial. $Bi_2Se_3$ nanowires are formed by stacking of nanoplatelets (Figure 1E,F), exhibiting rough surfaces along the wire direction. The size of the platelets is comparable to the size of the catalyst. The nanowire growth direction is determined by the way of stacking, either in parallel to $c$-axis (Figure 1E) or off $c$-axis (Figure 1F). These nanowires usually range from several hundred nanometers to 20 $\mu$m length, consisting of hundreds of nanoplatelets. $Bi_2Se_3$ nanoribbons, on the other hand, have flat surfaces and diverse morphologies. The thickness of these ribbons is roughly determined by the size of the catalyst, and the width of the ribbon varies from 50 nm to tens of micrometers. The growth of narrow ribbons (Figure 1G) is dominated by one-dimensional growth mode. The typical length of the narrow nanoribbons is several micrometers, but ultralong ribbons with lengths over 100 $\mu$m are often observed as well, as shown in Figure 1B. Wide nanoribbons, on the other hand, (Figure 1H) favor planar growth to form a sheetlike shape. Zig-zag (Figure 1I) and sawtooth (Figure 1J) nanoribbons are produced due to the repetitive switch of the growth direction. There are also a small amount of nanoplates (Figure 1K) grown on the substrate with the size of approximately hundreds of nanometers. The absence of the Au nanoparticle catalyst suggests vapor−solid (VS) growth mechanism for these nanoplates.

To obtain the crystal structure and chemical composition of the synthesized products, $Bi_2Se_3$ nanoribbons and nanowires are characterized by a 200 kV FEI Tecnai F20 transmission electron microscope (TEM), energy-dispersive X-ray spectroscopy (EDX) equipped in the TEM, and a Rigaku D/MAX-IIIC X-ray diffractometer (XRD). Figure 2A is a TEM image of a narrow nanoribbon with a similar morphology to that shown in Figure 1G. The corresponding selected-area electron diffraction (SAED) pattern (inset of Figure 2A) exhibits a clear hexagonally symmetric spots pattern, confirming the single crystalline nature. A high resolution TEM (HRTEM) image (inset in Figure 2A) shows hexagonal lattice fringes with a correct lattice spacing of 2.1 Å between (11-20) planes. SAED patterns and HRTEM images demonstrate that the nanoribbons grow along [11-20] direction, with (0001) facets as top and bottom surfaces and (1-100) facets as side surfaces. As a result, nanoribbons have a rectangular cross section. TEM studies on a wide nanoribbon (Figure 2B) reveal similar crystalline characteristics to the narrow nanoribbon (Figure 2A); it grows along [11-20] direction as well. The TEM data of a nanowire with rough surface are shown in Figure 2C. A spot SAED pattern on the wire demonstrates its single crystalline nature. From the low-magnification TEM image, the nanowire is composed of nanoplatelets arranged in parallel. For this nanowire in particular, the growth direction is close to [2 -1 -1 30]. The expitaxial growth of a nanowire on top of a nanoribbon (Figure 2D) is also observed. The SAED spot pattern and HRTEM image of the nanoribbon are shown in the inset. Small platelets in parallel to the nanoribbon pile together to form the nanowire. To reveal the chemical composition of the nanoribbons and nanowires, we acquired EDX spectra of the nanomaterials in TEM (Supporting Information S1−S3). Within experimental uncertainty limit of EDX analyses (∼1−2%), Bi/Se atomic ratio of the nanoribbons and nanowires are determined as 2:3, demonstrating the synthesized nanomaterials are indeed $Bi_2Se_3$. The large yield of nanomaterials allows us to characterize the growth product by powder X-ray diffraction (PXRD) directly on as-grown substrates (Supporting Information S4). Major diffraction peaks match well to that of the rhombohedral phase of $Bi_2Se_3$. No other Bi−Se compound crystalline phases are observed, which demonstrates the effective control of Bi/Se stoichiometry over the whole substrate by VLS growth mechanism.

To characterize the surface properties of the $Bi_2Se_3$ nanoribbons, we performed STM measurements on ribbons transferred to a metallic Au (111) substrate (preparation details in Supporting Information). All STM data shown were acquired in constant−current mode at 78 K in an ultrahigh vacuum chamber (pressure <$10^{-10}$ torr) with a sample bias of +0.5 V and tunneling currents between 20 and 50 pA. The STM (customized Omicrometer LT STM) has a scan range of ~2.2 $\mu$m while operating at 78 K. A composite low magnification STM topography image consisting of sixteen 2.2 $\mu$m scans is shown in Figure 3A with the color-scale representing the topographic height (details in Support Information). The composite image shows two $Bi_2Se_3$ nanoribbons, as well as a nanoplate lying on top of the ribbons (center of the image). Smaller scans allowed us to investigate the atomic-scale features of the nanoribbons. Figure 3B shows a topographic scan on top of a nanoribbon that exhibits individual terraces. A linecut taken across these terraces (Figure 3C) shows the height of two steps as 10.2 and 10.8 Å respectively, consistent with the $Bi_2Se_3$ quintuple layer thickness. This correspondence suggests a quintuple layer-by-layer growth process for $Bi_2Se_3$ nanoribbons. Figure 3D shows a 3D image of the region where the nanoribbons intersect the nanoplates. Here the orientation of the nanostructures' edges and the terrace edges are visible, occurring at multiples of 60° in accordance with the two-dimensional triangular lattice of the $Bi_2Se_3$ surface layer. We resolve the atomic lattice of the $Bi_2Se_3$ surface (Figure 3E) with a correct lattice constant of 4.04 ( 0.07 Å as determined by fast Fourier transformation (FFT) in Figure 3F. One interesting feature of the atomically resolved topography is that the data shows a honeycomb lattice (similar to that of graphene), rather than a purely triangular one. This is not expected, since the surface atomic layer of $Bi_2Se_3$ consists of a triangular lattice of Se atoms (see Figure 1A). One possible explanation for the honeycomb symmetry is that the STM tip may be strongly coupled to the Bi atoms, which compensates the reduction in tunnel current due to the greater distance between the tip and Bi atoms. As a result, the Bi atomic layer is also visible in the topography data, consequently creating the two-dimensional honeycomb lattice.

The morphology of $Bi_2Se_3$ nanoribbons allows facile fabrication of devices. After a nanoribbon sample is selected, multiple electrodes are patterned into a six terminal hall bar configuration, using a standard e-beam lithography technique and e-beam evaporation of Ti/Au (5 and 195 nm, respectively). A SEM image of a typical Hall bar device is shown in the inset of Figure 4. The hall bar electrode configuration allows us to measure the four-terminal resistance $R$ and hall resistance $R_H$ of the same ribbon. Transport measurements are performed in a Quantum Design PPMS-7 instrument with a base temperature of 2 K and a maximum magnetic field of 9 T. Four-terminal resistance is obtained by flowing a known current through the two outer contacts and monitoring the voltage drop between the two inner contacts. Temperature dependent, four-terminal resistance from room temperature down to 2 K is shown in Figure 4A. The resistance decreases with the temperature decrease and reaches a saturation minimum value below 20 K, in agreement with a typical behavior of a heavily doped semiconductor, which resulted from the small bandgap of $Bi_2Se_3$ and the residual doping from intrinsic defects such as Se vacancies.[4,22,23] To confirm the doped nature of as-grown $Bi_2Se_3$ nanoribbons, the Hall resistance is measured to directly probe the carrier density. In a Hall measurement, a known current is driven by outermost contacts, and the potential difference between opposite electrodes perpendicular to the current flow is recorded. By measuring over 10 hall bar devices, the magnetic field ($B$) dependence of Hall resistance ($R_H$) generally shows two types of behaviors. The $R_H-B$ curve with a small slope initially has a linear dependence with the magnetic field up to 9 T, corresponding to a high carrier area density of $1.5 \times 10^{14}$ cm$^{-2}$.

The $R_H-B$ curve with a larger slope in the low field region, on the other hand, deviates from the linear behavior observed at high field, indicating the existence of different types of carriers.[24] The corresponding carrier area density at the low field is $2.6 \times 10^{13}$ cm$^{-2}$. The nonlinear dependence of $R_H$ on the $B$ field may be ascribed to the presence of the surface states, previously masked by excessive bulk carriers in a high density sample. In both Hall traces, the sign of the Hall slope demonstrates that as-grown $Bi_2Se_3$ nanoribbons are a doped n-type semiconductor. Hall measurement thus indicates synthesized $Bi_2Se_3$ nanoribbons have certain variations in the Se vacancy concentration. The approach to effectively reduce Se vacancies in the as-grown nanoribbons is still under study. Despite the high bulk carrier concentrations, nanoribbons with a large surface-to-volume ratio can still manifest the contribution of surface conducting states, and open up exciting opportunities for future surface topological state property investigation.[25]

In summary, synthesis of topological insulator material $Bi_2Se_3$ nanowires and nanoribbons is achieved based on the VLS growth mechanism. The synthesized $Bi_2Se_3$ nanomaterial exhibits diverse morphologies, controllable by growth conditions. Nanoribbons have atomically flat surfaces with 1 nm quintuple layers as step edges, as ideal candidates for scanning probe measurement and devices patterning. Preliminary device measurements show great promise toward understanding topological surface states by electron transport. Topological insulator nanowires serve as exciting candidates for building next generation spintronics devices.

**Acknowledgment.** Y.C. acknowledges the support from the King Abdullah University of Science and Technology (KAUST) Investigator Award (No. KUS-l1-001-12). STM work (J.C.R. and H.C.M.) and part of transport measurements (Z.-X. S.) were supported by the Department of Energy, Office of Basic Energy Sciences, Division of Materials Sciences and Engineering, under contract DE-AC02-76SSF00515. K.L. acknowledges the KAUST Postdoctoral Fellowship support (No. KUS-F1-033-02).

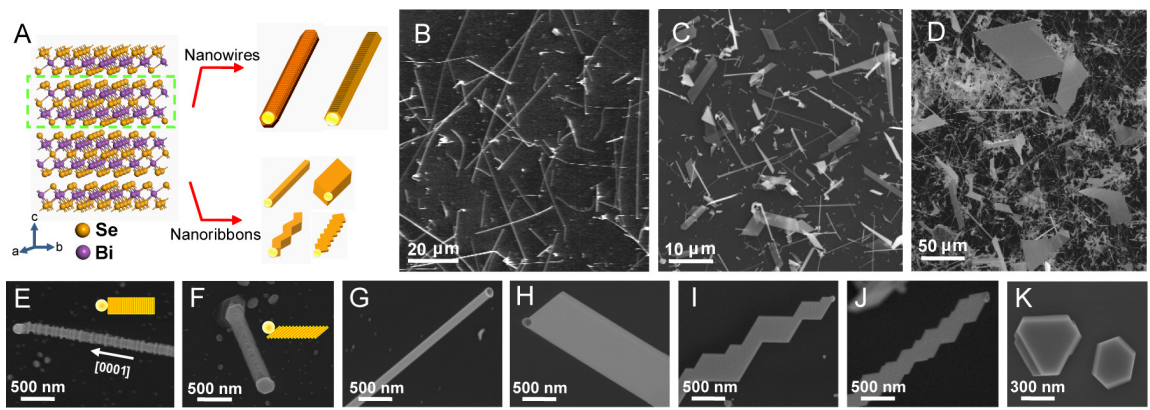

FIGURE 1. SEM images of synthesized Bi$_2$Se$_3$ nanomaterials. (A) Crystal structure of Bi$_2$Se$_3$ and different morphologies of synthesized nanomaterials. Dashed green lines indicate the Se−Bi Se−Bi−Se quintuple layer with the thickness of ∼1 nm. (B) Nanowires grown on a warm substrate (∼8 cm from the center) by a 530 °C synthesis with 30 sccm Ar for 1.5 h. (C) Bi$_2$Se$_3$ nanoribbons grown on a cool substrate (∼12 cm from the center) by a 530 °C synthesis with 30 sccm Ar for 1.5 h. (D) Huge nanoribbons grown at 580 °C (the center of the furnace) with 120 sccm Ar for 5 h. (E) Nanowire grown along *c*-axis. (F) Nanowire grown off *c*-axis. (G) Quasi-1D narrow nanoribbon. (H) Sheetlike wide nanoribbon. (I) Zig-zag nanoribbon. (J) Sawtooth nanoribbon. The presence of gold nanoparticles at the end of individual nanostructure indicates VLS growth mechanism in (E−J). (K) Nanoplates grown on the substrate by VS mechanism.

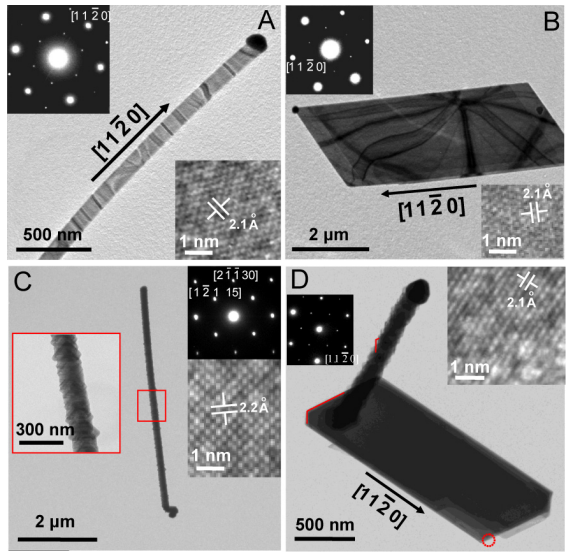

FIGURE 2. TEM images of Bi$_2$Se$_3$ nanowires and nanoribbons with corresponding HRTEM images and SAED patterns as the insets. (A,B) TEM studies on a Bi$_2$Se$_3$ narrow and wide nanoribbon respectively. Nanoribbons grow along [11-20] direction. (C) TEM data of a nanowire with the growth direction close to [2 -1 -1 30]. (D) Epitaxial growth of a nanowire on top of a wide nanoribbon. HRTEM and SEAD images are obtained from the area indicated by the dashed circle. Edges of the nanoribbon and a platelet of the nanowire are highlighted with a red line to show the parallel arrangement.

height is consistent with a single quintuple layer thickness. (D) 3D topograph of nanoribbons and sheets. Arrows indicate the orientation of the triangular lattice vectors. (E) Topography of honeycomb atomic lattice. Arrows indicate the length and orientation of lattice vectors. Orange (magenta) dots show presumed positions of Se (Bi) atoms, which form the surface (subsurface) layer. (F) Magnitude of the Fourier transform of the atomically resolved topograph. Bright peaks correspond to the atomic lattice. The 6-fold symmetry of the reciprocal lattice has been exploited to symmetrize the image.

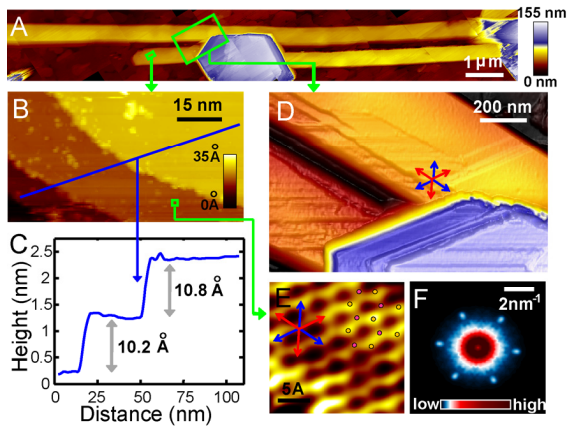

FIGURE 3. STM study of Bi$_2$Se$_3$ nanoribbons. (A) Composite STM topography of two nanoribbons and a nanoplate. Green boxes indicate approximate locations of smaller scans depicted in this figure. (B) Topography showing terraces on the nanoribbon surface. (C) Line-cut showing the ribbon height across three steps (blue line in (B)); step

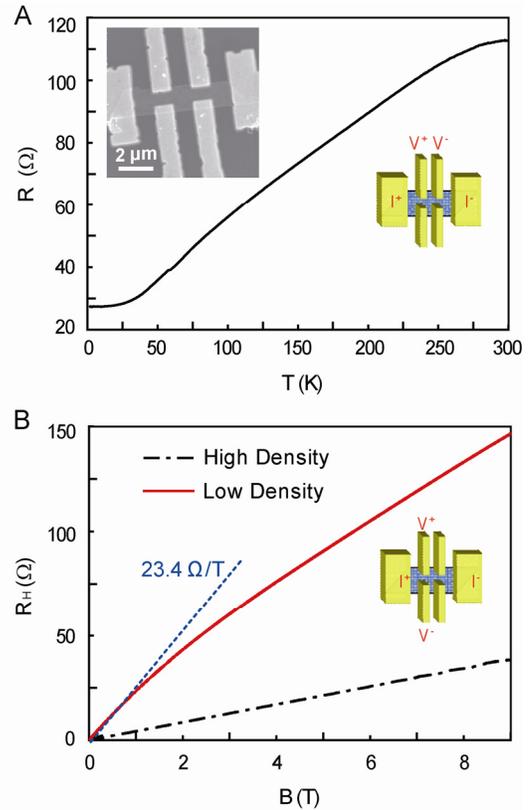

FIGURE 4. Electrical measurement of a nanoribbon device. (A) Temperature dependence of four-terminal resistance from room temperature to 2 K. Insets are a SEM image of the Hall bar device and the measurement setting. (B) Hall traces of a representative Hall bar devices (inset) measured at 2 K with the measurement setting shown in the inset.